\documentclass[12pt,preprint,longabstract]{aastex}
\usepackage{graphicx}
\usepackage{fullpage}
\usepackage{amsmath}
\usepackage{enumerate}
\usepackage{txfonts,epsfig,graphicx,float}
\usepackage{subfigure}
\usepackage{epstopdf}

\shorttitle{Dynamical model for spindown of solar-type stars}
\shortauthors{Sood, A., Kim, E. and Hollerbach, R.}

\begin{document}

\title{Dynamical model for spindown of solar-type stars}

\author{Aditi Sood\altaffilmark{1}, Eun-jin Kim\altaffilmark{1}}

\and

\author{Rainer Hollerbach\altaffilmark{2}}

\altaffiltext{1}{School of Mathematics and Statistics, University of Sheffield, Sheffield S3 7RH, United Kingdom}
\altaffiltext{2}{Department of Applied Mathematics, University of Leeds, Leeds LS2 9JT, United Kingdom}

\begin{abstract}
Since their formation, stars slow down their rotation rates by the removal of angular momentum from their surfaces, e.g. via stellar winds. Explaining how this rotation of solar-type stars evolves in time is an interesting but difficult problem in astrophysics in present times. Despite the complexity of the processes involved, a traditional model, where the removal of angular momentum loss by magnetic fields is prescribed, has provided a useful framework to understand observational relations between stellar rotation and age and magnetic field strength. Here, for the first time, a spindown model is proposed where loss of angular momentum by magnetic fields is evolved dynamically, instead of being kinematically prescribed. To this end, we evolve the stellar rotation and magnetic field simultaneously over stellar evolution time by extending our previous work on a dynamo model which incorporates the nonlinear feedback mechanisms on rotation and magnetic fields. We show that our extended model reproduces key observations and is capable of explaining the presence of the two branches of (fast and slow rotating) stars which have different relations between rotation rate $\Omega$ vs.
time (age), magnetic field strength $|B|$ vs. rotation rate, and frequency of magnetic field $\omega_{cyc}$ vs. rotation rate.
 For fast rotating stars we find: (i) there is an exponential spindown $\Omega \propto e^{-1.35t}$, with $t$ measured in Gyrs, (ii) magnetic activity saturates for higher rotation rate, (iii) $\omega_{cyc} \propto \Omega^{0.83}$. For slow rotating stars we obtain: (i) a power law spindown $\Omega \propto t^{-0.52}$, (ii) magnetic activity scales roughly linearly with rotation rate, (iii) $\omega_{cyc} \propto \Omega^{1.16}$. The results obtained from our investigations are in good agreement with observations. The Vaughan-Preston gap is consistently explained in our model by the shortest spindown timescale in this transition from  fast to slow rotators. Our results highlight the importance of self-regulation of magnetic fields and rotation by direct and indirect interactions involving nonlinear feedback in stellar evolution.
\end{abstract}


\section{Introduction}
Spindown of stars is one of the most debated and interesting issues in
astrophysics. Stellar rotation rate is the key parameter which is believed to
affect the spindown process. Spindown is not only influenced by stellar
properties such as mass, radius and age, but also depends upon the evolution of
stellar magnetic fields and their interaction with the stellar atmosphere (Scholz,
2008). Since their formation from interstellar clouds, which involves various
internal changes, stars undergo rotational evolution in different stages {(Keppens et al. 1995, Tassoul 2000)}, briefly
summarized in the following. {During early pre-main sequence evolution, the
contraction that occurs in the star along with other various internal structural changes lead it to spin-up.
Also, owing to diverse internal changes a radiative core develops which rotates
faster than the convective envelope. Coupling between radiative core and
convective envelope {should be strong enough for the angular momentum to be constantly transferred from core to envelope. This persistent supply of angular momentum from core to envelope reduces the amount of differential
rotation produced in the star.}
By the time the star reaches late pre-main sequence or early main sequence, rotational evolution is modified by the stellar wind.
Angular momentum loss via stellar wind gradually decelerates and stops the spin-up of
convective envelope towards the end of late pre-main sequence phase and causes a fast spindown of convective envelope on the main sequence. Timescale at which decoupling of core and envelope occurs is
observed to be very rapid (Keppens et al. 1995)}. With increasing rotation, timescale for angular momentum
loss through stellar wind decreases and affects the magnetic field strength.
Consequently, for rapidly rotating stars, the magnetic field strength does not
increase beyond a critical value at a certain rotation rate and instead becomes
independent of rotation no matter how fast the star is rotating. When convection
zone spins down towards the end of pre-main sequence, magnetic field strength is
believed to scale linearly with rotation rate in case of slow rotating stars.\\

\indent Based upon the whole spindown process stars are often classified into two groups: fast and slow rotating
 (Saar \& Brandenburg 1998, Brandenburg et al. 1999, Barnes 2003, Pizzolato et al. 2003, Mamajek \& Hillenbrand 2008, Wright et al. 2011, Vidotto et al. 2014). The existence of
two branches of stars, exhibiting different dependence of {cyclic variation of stellar magnetic activity known as} cycle period $P_{cyc}$ on rotation period $P_{rot}$, was confirmed by Saar and Brandenburg (1998)
and later by Brandenburg et al. (1999). We note that a relationship between cycle period and rotation period was first established by Noyes et al. (1984) as $P_{cyc}\propto P_{rot}^{n}$ with  $n=1.25\pm0.5$.  Brandenburg et al. showed all young, active and fast rotating stars lie on
one branch namely active branch (A) with scaling exponent $n=0.80$, while all old, inactive and slow rotating stars lie on other branch namely inactive branch (I) with scaling exponent $n=1.15$ ({Saar \& Brandenburg 2001}, Charbonneau \& Saar 2001). Furthermore, stars on $A$ branch experience rapid spindown for which rotation rate $\Omega$ is related to time/{age} with an exponential law given as $\Omega \propto e^{mt}$, where $m$ is a negative constant, and in this case magnetic activity is found to be saturated, {that is, magnetic activity becomes independent of rotation rate for rapidly rotating stars}. Stars on $I$ branch
undergo a very slow spindown with a power law dependence as $\Omega \propto t^{-1/2}$, known as power law spindown
(Skumanich 1972), and in this case magnetic activity is thought to scale linearly with rotation rate. {The relationship between magnetic activity and rotation rate is important to understanding the physical process responsible for spindown of a star and was first determined by Pallavicini et al. (1981), while Micela et al. (1985) observed that this relationship does not hold for rapidly rotating stars. We note that the regime where magnetic activity increases linearly with rotation rate is termed as `unsaturated (non-saturated) regime' while the regime where magnetic activity becomes independent of rotation rate is termed as `saturated regime' in observational studies (e.g. Pizzolato et al. 2003, Mamajek \& Hillenbrand 2008, Wright et al. 2011, Vidotto et al. 2014).}\\

\indent One of the challenging problems in explaining spindown is the existence of a gap between the two branches of stars. During the spindown, the star suddenly jumps from $A$ to $I$ branch, creating a gap between the two
 branches where stars are sparsely populated. This gap was first observed by Vaughan and Preston (1980) and is now known as the V-P gap. Various mechanisms have so far been proposed for this gap, but the underlying physics is still an open question. Some of the
 previous suggestions are as follows. Durney et al. (1981) advocated a change in magnetic field morphology from complex to simple at the time when rotation decreases to a certain value. Saar (2002) proposed that the existence of two distinct branches of stars could be due to the changes in
 differential rotation, $\alpha$-effect and meridional flow speed (which is proportional to $\Omega$ in case of flux transport models, e.g. see Dikpati \&
 Charbonneau 1999) with stellar rotation rate. Barnes (2003) studied period-color-diagrams of open clusters and mentioned that the transition from convective (fast rotators) to interface sequence (slow rotators) is due to the shear produced
 during decoupling of core and envelope. This shear gives rise to large-scale magnetic fields, and recoupling of the core and convection zone shifts the star from
 convective sequence to interface sequence. Structural changes in large-scale magnetic fields (Donati et al. 2006), change in dynamo action (B\"ohm-Vitense
 2007) and manifestation of different dynamos for different stars (Wright et al. 2011) were also proposed as possible reasons for the V-P gap.\\

\indent
 Given the complexity of the spindown problem, which depends
upon various parameters such as rotation rate, evolution of magnetic fields and differential rotation, it is not possible to study a full magnetohydrodynamic model over the entire spindown timescales (e.g. from $10^7 -10^9$ yrs). Therefore, various simplified models have been utilized to understand stellar evolution
{ (e.g. Weber \& Davis 1967, Mestel 1968, Mestel \& Spruit 1987, Kawaler 1988, Matt et al 2012, Matt et al 2015, Johnson et al 2015, Cranmer \& Saar 2011, Cohen et al 2009, Garraffo et al 2015)}.
One such model is double zone model (DZM) which is based upon the stellar wind torque law (Weber \& Davis 1967; Mestel 1968; Belcher \& MacGregor 1976; Kawaler 1988). The main feature of this model is the bifurcated expression considered for the torque acting on the star  (depending on the critical rotation rate)
due to its magnetised stellar wind.
MacGregor and Brenner (1991) used this DZM model for coupled (ordinary differential) equations for the rotation rates of the stellar envelope and radiative core, where the angular momentum loss is prescribed according to the relation between rotation and magnetic field strength. To understand the distribution of stellar rotation at different ages,
Keppens et al. (1995) extended this parameterized model to describe the evolution of a single star by taking into account
angular momentum exchange, moment of inertia evolution and torque exerted on core and envelope due to which angular momentum changes. Since then, this model was extended by considering different initial conditions and tested against various observations in the spindown process (Krishnamurthi et al. 1997; Irwin \& Bouvier 2009; Denissenkov et al. 2010; Kim \& Leprovost 2010;
Epstein \& Pinsonneault 2014; Reiners \& Mohanty 2012; Spada et al. 2011; Gallet \& Bouvier 2013). Apart from DZM there are other models such as symmetrical empirical model (SEM) (Barnes 2010; Barnes \& Kim 2010) and  metastable dynamo model (MDM) (Brown 2014).
{{
Both SEM and MDM utilise observational data of two different sequences of stars to fine-tune their models and thus are descriptive rather than explanatory models.
Specifically, SEM uses different period-evolution of the two sequences (for active and inactive stars) depending on whether the rotation rate is above/below the critical value and fits the parameters from period-color diagrams by obtaining a best fit to the observational data. Unlike SEM, MDM uses one function for all rotation rates but two different coupling constants. By fine-tuning the values of these two coupling constants and the probability for the transition from small to large couplings, MDM improves the agreement with observations over SEM. Although it is yet empirical, MDM is remarkable in introducing into a spin-down  model
a threshold-like behaviour with different coupling constants and their probabilistic nature.
Possible mechanisms for these different coupling constants was later provided, e.g. by evoking the change in magnetic complexity
(Reville et al 2015, Garraffo et al 2015).
}}
Recently, Matt et al. (2015) proposed a stellar wind torque model (SWTM) which reproduces the shape of upper envelope and lower envelope that corresponds to the transition region between saturated and unsaturated regimes by explaining the mass-dependence of stellar magnetic and wind properties.\\

\indent In this paper, we for the first time propose a dynamical model of spindown where the loss of angular momentum by magnetic field is dynamically treated, instead of being kinematically prescribed. To this end, we evolve the stellar rotation and magnetic field simultaneously over the stellar evolution time by
 extending our previous work (Sood \& Kim 2013, 2014) which incorporates the nonlinear feedback mechanisms on rotation and magnetic fields via $\alpha$-quenching and magnetic flux losses as well as mean and fluctuating rotation.
We note that Sood \& Kim (2013, 2014) have demonstrated that nonlinear feedback plays a vital role in the generation and destruction of magnetic fields
as well as self-regulation of the dynamo. In particular, it was found that a dynamic balance is required not only in the generation and destruction of magnetic fields, but also in the fluctuating and mean differential rotation for the working of dynamo near marginal stability; their results were consistent with observations
 such as linear increase in cycle frequency of magnetic field with moderate rotation rates, levelling off of magnetic field strength with sufficiently large rotation rates, and quenching of shear. We extend this model to simultaneously evolve rotation and magnetic fields over the spindown timescale of a star, since their dynamics are closely linked through angular momentum loss and dynamo. That is, the angular momentum loss responsible for the spindown of a star depends upon magnetic fields while magnetic fields are affected by
 rotation rates. We show that this model has the capability of explaining the existence of the two branches of stars, different rotation rate dependence of cycle frequency of magnetic fields for these two branches, and the gap between the two branches, reproducing the main observations. By extending our previous work, our model is designed in such a way that it has {essential} ingredients mentioned above to explain the complex process of spindown of solar-type stars and highlight the importance of nonlinear feedback in this process.\\

\section{Model}
We propose a dynamical model for the evolution of rotation rate and magnetic field in spindown by extending a previous nonlinear dynamo model (Sood \& Kim, 2013, 2014; Weiss et al. 1984). In particular, Sood \& Kim (2013, 2014) incorporated various nonlinear transport coefficients such as $\alpha$-quenching and flux losses and took the control parameter $D$ known as the dynamo number to scale with rotation rate as $D\propto\Omega^2$. The model equations in dimensionless form are given as
\begin{eqnarray}
&&\dot{A} = \frac{2DB}{1+\kappa(|B|^2)} -[1+\lambda_{1}(|B|^2)]A, \\
&&\dot{B} = i(1 + w_{0})A - \frac{1}{2}iA^{*}w -[1+\lambda_{2}(|B|^2)]B, \\
&&\dot{w_{0}} = \frac{1}{2}i(A^{*}B - AB^{*}) - \nu_{0}w_{0}.\\
&&\dot{w} = -iAB - \nu w.
\end{eqnarray}
Here, poloidal magnetic field is represented by $A$, toroidal magnetic field is given by $B$, $w_{0}$ is the mean differential rotation, and $w$ is the fluctuating differential rotation; $A$, $B$ and $w$ are complex variables whereas $w_{0}$ is real. We note that $w_{0}$ and $w$ have zero and twice the frequency of $A$ and $B$, respectively. The complex conjugates of $A$ and $B$ are denoted by $A^{*}$ and $B^{*}$, respectively. {In this model,  poloidal magnetic field $A$ is generated by toroidal magnetic field $B$ (e.g. $\alpha$-effect through helicity) which is assumed to be proportional to rotation rate $\Omega$ (see Eq. 1). Equation 2 represents the generation of toroidal magnetic field $B$ by poloidal magnetic field $A$, where the quenching of $\Omega$-effect is incorporated by total shear $1+w_{0}$. The differential
rotation is inhibited by the tension in the magnetic field
lines via Lorentz force and causes the quenching of $\Omega$-effect. Due to back-reaction, the total shear is reduced from 1 to $1+w_{0} < 1$ as $w_{0}$ is always negative and is given by $1+w_{0} = {\Delta\Omega}/{\Omega}$. Generation of mean differential rotation $w_{0}$ and fluctuating differential rotation $w$ is represented by Eq. 3 and Eq. 4, respectively.}
$\nu_{0}$ and $\nu$ represent viscosity of mean differential rotation and fluctuating differential rotation, respectively; $\kappa$, $\lambda_{1}$ and $\lambda_{2}$ are constant parameters which represent the strength of nonlinear feedback due to the Lorentz force by magnetic field and enhanced magnetic dissipation (e.g. magnetic flux loss). {In particular, $\kappa$ represents the efficiency of the quenching of $\alpha$-effect while $\lambda_{1}$ and $\lambda_{2}$ represent the efficiency in the poloidal and toroidal magnetic flux losses, respectively} (see Sood \& Kim 2013 for full details).\\

\indent
To understand the evolution of rotation rate and magnetic field in spindown of solar-type stars, we extend this model by upgrading $\Omega$ from a kinematically prescribed to a dynamic variable. To this end, we first replace $D$ by the square of time dependent rotation rate {$\Omega$}(t) in Eq. 1:
\begin{eqnarray}
&&\dot{A} = \frac{2\Omega^2B}{1+\kappa(|B|^2)} -[1+\lambda_{1}(|B|^2)]A,
\end{eqnarray}
{where $\Omega$ is real}.
Second, we { need to} include the additional equation for the evolution of $\Omega(t)$ to model the spindown of a star by the  loss of angular momentum due to magnetic fields.
{
While  the latter depends on many factors such as the mass flux and geometry and complexity of magnetic fields (e.g. Garraffo et al 2016) such as the Alfven radius over which it acts as a rotational brake and the latitude at which the mass release happen, for simplicity, we incorporate their overall effects in our dynamical model by
the ansatz that a decay rate of $\Omega$ is proportional to
the strength of magnetic fields as $\varepsilon_{1}|B|^2 + \varepsilon_{2}\frac{|A|^2}{\Omega}$ with the two tunable parameters $\epsilon_{1}$ and $\epsilon_{2}$. Here, $|B|$ represents the strength of toroidal magnetic field and $|A|\over \Omega$ is the strength of
 poloidal magnetic field in physical units due to our non-dimensionalisation (see Sood \& Kim 2013).
Our {empirical model}  is thus described by the following equation for $\Omega$:
}
\begin{eqnarray}
&&\dot{\Omega} = -\varepsilon_{1}|B|^2\Omega - \varepsilon_{2}\frac{|A|^2}{\Omega^2}\Omega.
\label{awesome}
\end{eqnarray}
Eq. 6 represents the overall spindown of the star as a whole due to the loss of angular momentum through magnetic fields. Constant parameters $\varepsilon_{1}$ and $\varepsilon_{2}$ represent the efficiency of angular momentum loss via toroidal and poloidal magnetic fields, respectively,
{
which are taken to be independent in general, given the uncertainty in precise role of poloidal and toroidal magnetic fields in spin-down.  Eq. \ref{awesome} is motivated to capture the key feature of the previous model (e.g. DZM) where the dependence of the angular momentum loss  on $\Omega$ is roughly  proportional to $\Omega^3$ for slowly rotating stars (below the critical
rotation rate) to $\Omega$ for fast rotating stars (above the critical rotation rate), respectively.
Specifically, for fast rotating stars with the rotation rate  above the critical value,
$|B|$ and $|A|$ become independent of $\Omega$, Eq. \ref{awesome} reducing to $\dot{\Omega}  \propto - \Omega$,
resulting in the exponential decay of $\Omega$ in time.
On the other hand, for slow rotating stars, $\dot{\Omega}  \sim - \Omega^3$ would be reproduced
should magnetic field increase linearly with $\Omega$ as $|B|, |A| \sim \Omega$ (see \S 3.2 for the scaling relation).
}
 To summarize, our extended model consists of Eqs. 2-4
 and 5-6, where Eqs. 2-4 are the same as in our previous model, Eq. 5 is the modified form of Eq. 1, and Eq. 6 is a new equation to model the time-evolution of $\Omega$. \\

\indent
This system is investigated taking $\nu =0.5$, $\nu_{0}=35.0$, $\kappa=0.025$,
$\lambda_{1,2} = 1.125$ and $\varepsilon_{1,2} = 3.5\cdot10^{-5}$. The parameters
$\nu$, $\nu_0$, $\kappa$ and $\lambda_{1,2}$ are much the same as in our
previous work (Sood \& Kim 2013, 2014). {As can be seen from Eq. 6,
the two new parameters $\varepsilon_{1,2}$ control the rate of the spindown
process. The value $3.5\cdot10^{-5}$ was chosen as it yields an overall
spindown timescale of several Gyrs; larger (smaller) values of $\varepsilon_{1,2}$ were also investigated, and yielded qualitatively the same dynamics, simply occurring on shorter (longer) timescales.
{ In particular,  we have checked that qualitatively similar results are obtained in the limiting cases where $\epsilon_{1}=0$
or $\epsilon_{2}=0$.
}
Correspondingly, the dimensionless time scales such that the largely completed spindown process translates to the
present-day age of the Sun of 4.5 Gyrs.}

{To model the spindown process, we take the initial value of $\Omega$ to
be 30, corresponding to thirty times the present-day solar rotation, which is
$\Omega=1$ in our non-dimensionalisation. An initial value of $\Omega=30$ is
intended to model the rotation rate of young stars at an age of around
$\sim10^7$ years.} In contrast to $\Omega$, which can only decrease monotonically
according to Eq. 6, the initial conditions of the other four variables are not
important, as they can increase as well as decrease, and turn out to settle in
to statistically stationary states on comparatively rapid timescales; that is,
transients depending on initial conditions of these quantities quickly vanish,
 and the subsequent evolution depends only on the initial value chosen for $\Omega$. Finally, note that because $\Omega$ is monotonically decreasing in time, we can effectively invert the relationship $\Omega(t)$ as $t(\Omega)$, and therefore consider all the other variables as functions of $\Omega$ rather than $t$.

\section{Results}

\subsection{$\Omega$ versus age relationship}

Fig. 1 shows the relationship between $\Omega$ and $t$. A sharp decrease in
$\Omega$ can be seen for earlier times, which slows down as age starts increasing. In the left panel of Fig. 2 we fit this curve using an exponential law, that is, $\Omega \propto e^{mt}$.
 The best fit, for stars with rotation periods in the range
$1\leq P_{rot}\leq 3$, has $m=-1.35$, corresponding to an $e$-folding time of
0.74 Gyrs. In the right panel of Fig. 2 we fit this curve using power laws, that is, $\Omega \propto t^{n}$.
For larger times we get power law scalings which vary gradually for different rotation rates, that is,  { $n$ becomes smaller for smaller rotation as observed
in MacGregor and Brenner (1991)}. For larger age (slower rotation rates), the power law exponent $n$ is found to be around $-0.52$ for stars
with rotation periods in the range $23\leq P_{rot} < 25.65$. For different rotation rates we summarise the scalings in Table 1.\\
\begin{figure}[ht!]
\epsscale{1.2}
\plottwo{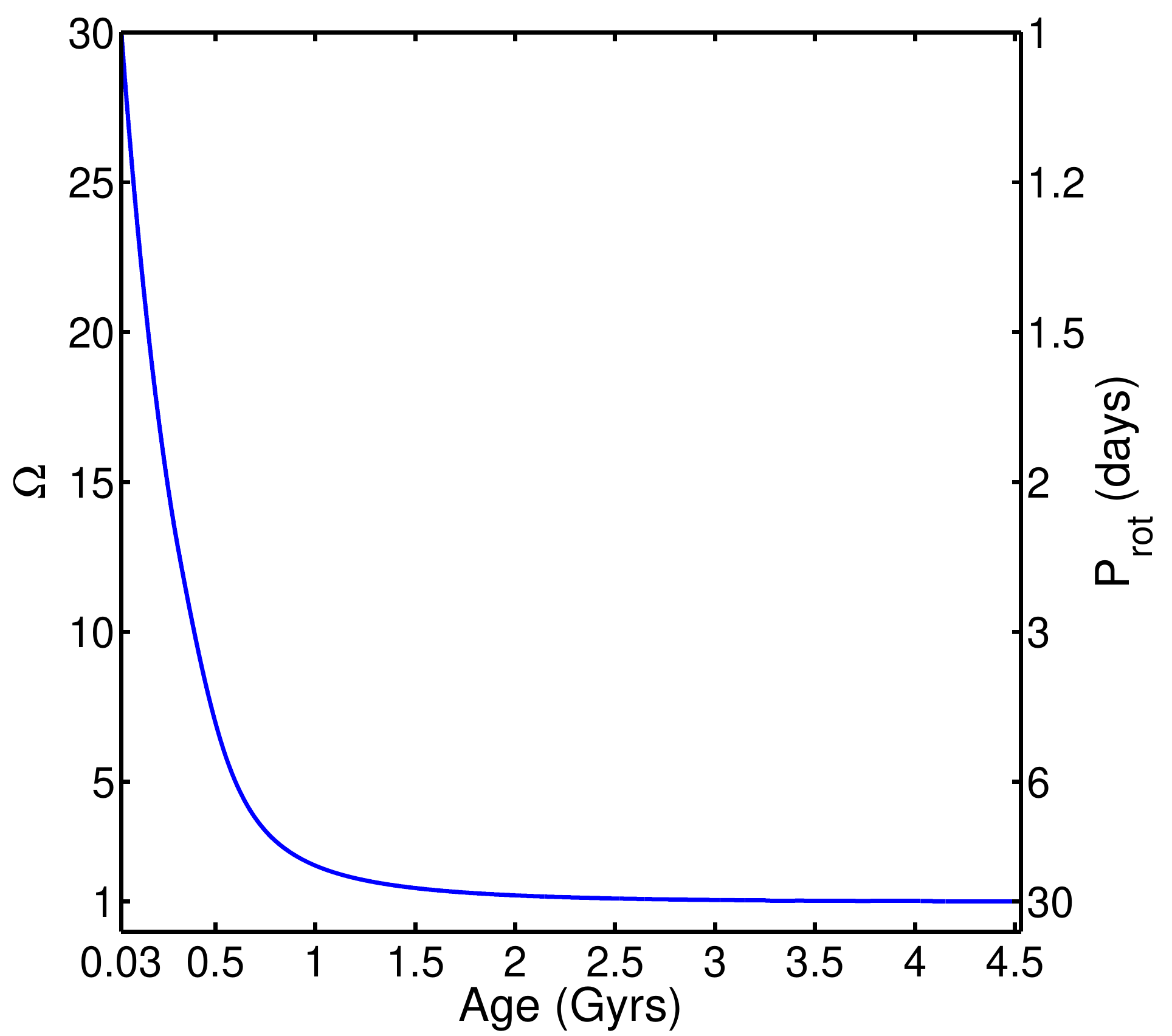}{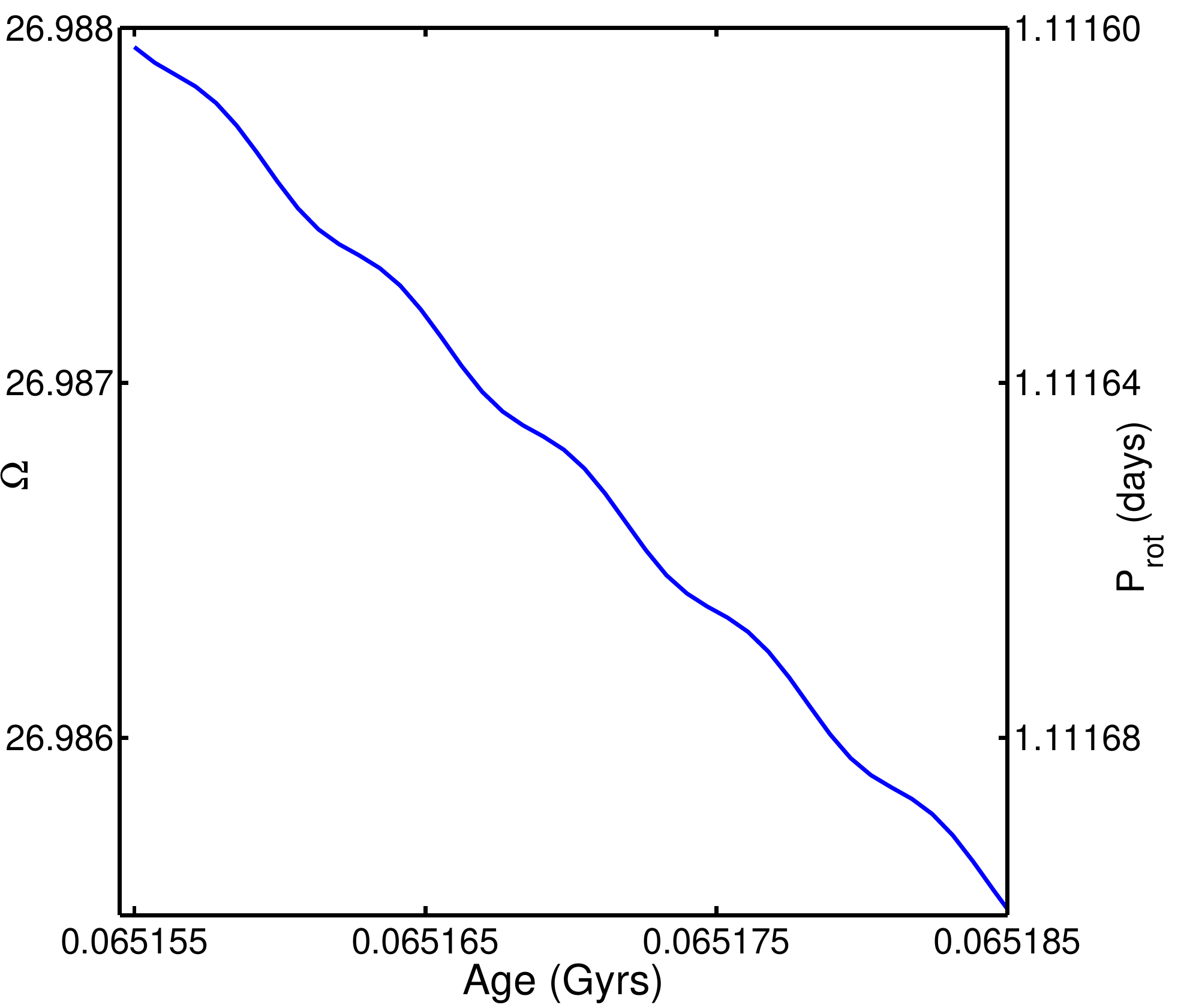}
\caption{The left panel shows $\Omega$ as a function of age over the full range {[0.03, 4.53] Gyrs}.  {As noted in section 3, we scale our dimensionless time in physical units by the age of the present-day Sun of 4.5 Gyrs, while our dimensionless rotation, represented on the vertical axis, is scaled with thirty times solar rotation to obtain rotation period ($P_{rot}$) in days, which is depicted on the right side of the $y$-axis.} The right panel shows a zoomed-in view for {age =[0.065155, 0.065185] Gyrs}, and reveals the presence of fluctuations superimposed on the general spindown trend.\label{fig1}}
\end{figure}
\begin{figure}[ht!]
\epsscale{1.0}
\plottwo{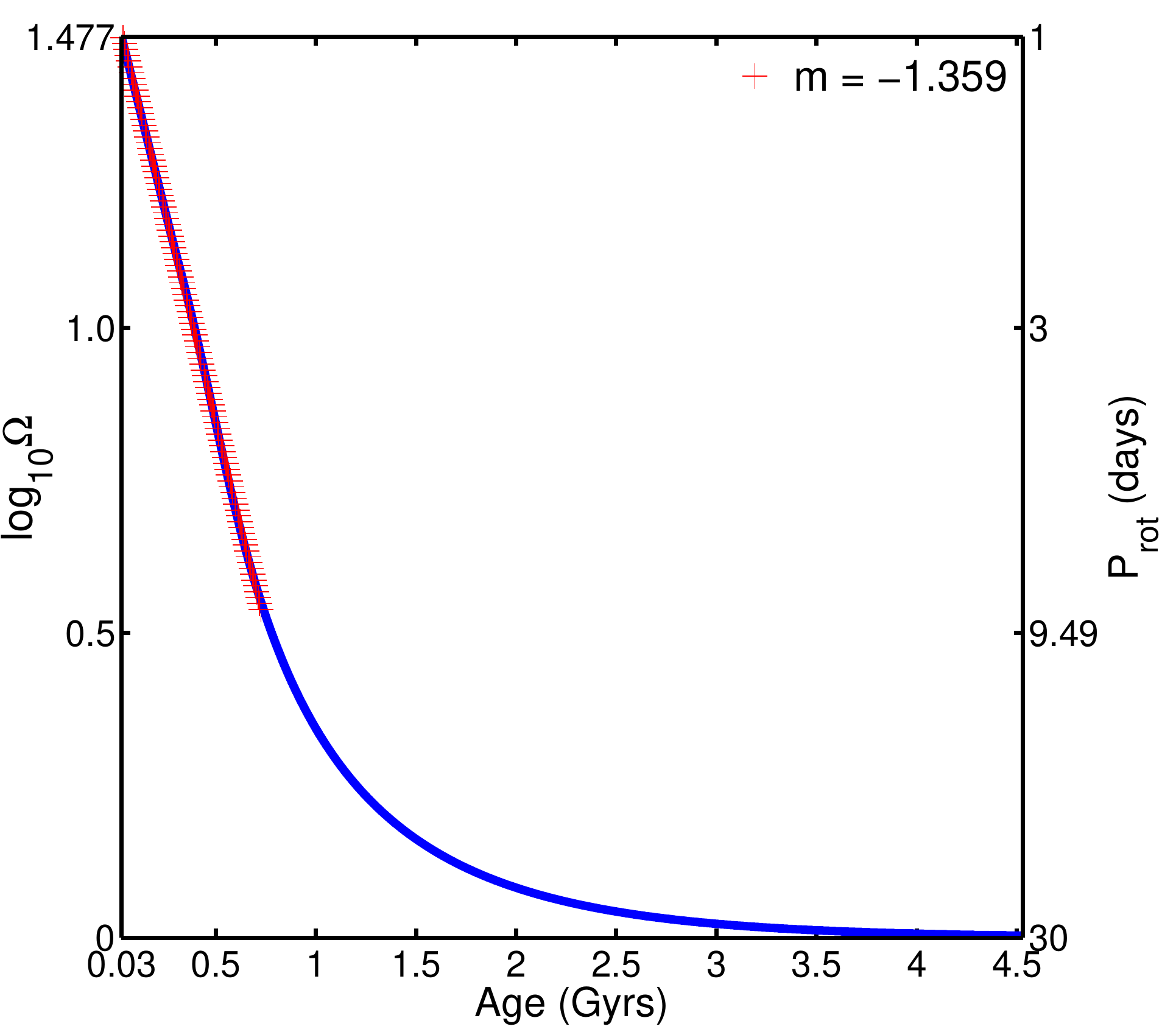}{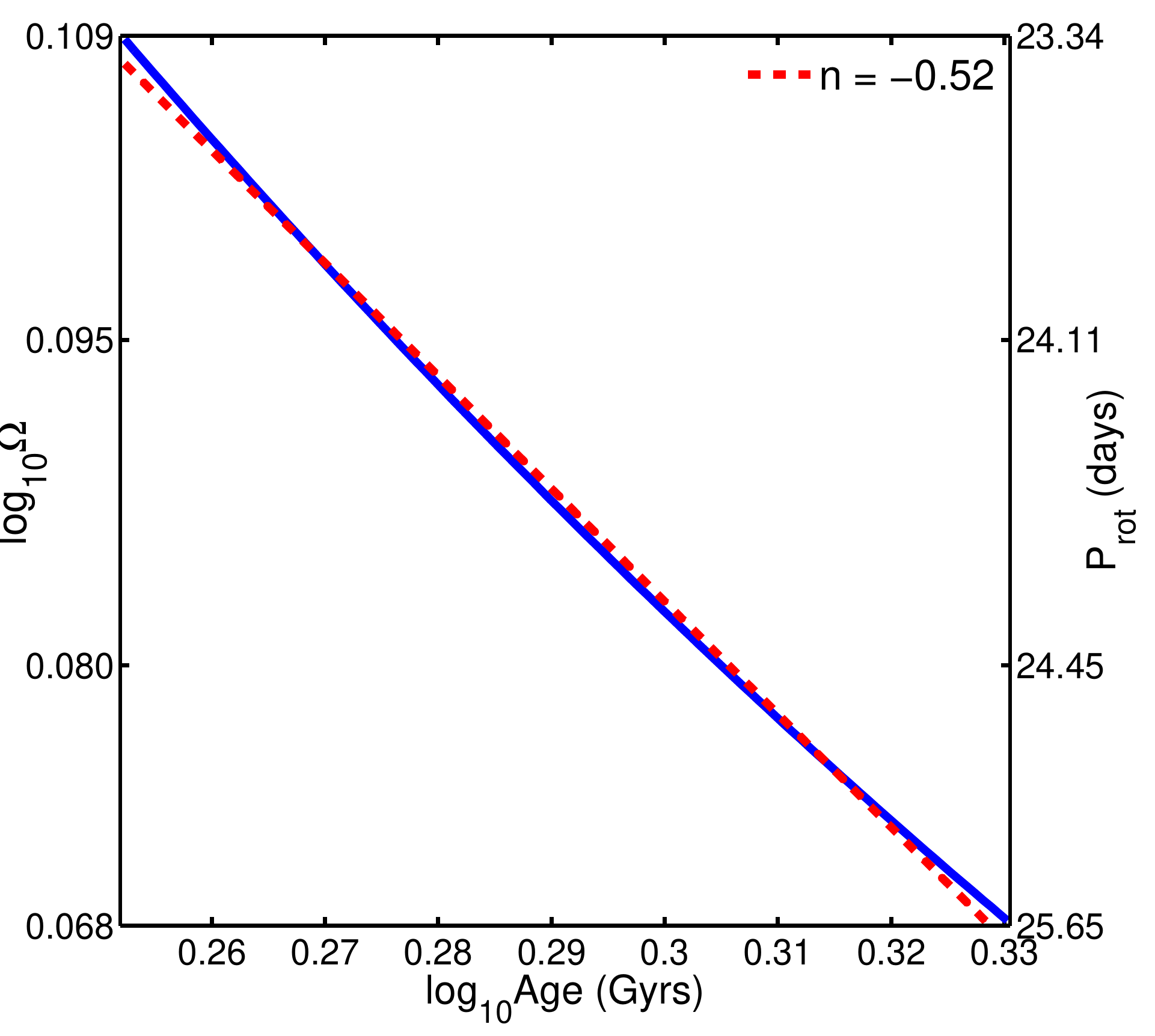}
\caption{Left panel shows exponential spindown with $\Omega\propto exp^{{-1.35}t}$ for ages {$\in[0.03,0.7325]$ Gyrs} depicted in red while the blue color represents the trend in semi-logarithmic scale for {age $\in[0.03,4.53]$ Gyrs}. Right panel shows the power law spindown,
$\Omega\propto t^{n}$, with scaling exponent $-0.52$ for solar-type stars.
{
A gradual decrease in $|n|$ suggests a drop in the efficiency of angular momentum loss, which seems to align with
the suggestion for the reduction in the efficiency of magnetic braking from recent observations from the Kepler space telescope
(e.g. Garraffo et al. 2016).
}
\label{fig2}}
\end{figure}
 \begin{table}[ht!]
\caption{Power law exponent $n$ for stars with different rotation period in days }              
\label{table:1}      
\centering
  \begin{tabular}{c  c c}
    \hline
       \hline

     $n$  &$\Omega$ & $P_{rot}(days)$ \\ \hline
     -1.38&$\Omega \in [3.5,1.99]$  & 8.57- 15 \\
     -0.97&$\Omega \in [1.99,1.50]$ & 15 - 20 \\
     -0.70&$\Omega \in [1.50,1.28]$ & 20 - 23.34\\
     -0.52&$\Omega \in [1.28,1.17]$ & 23.34 - 25.65\\

    \hline
  \end{tabular}

\end{table}

 \subsection{$|B|$ versus $\Omega$ relationship}

 {Magnetic field strength $|B|$ is shown as a function of rotation rate in Fig. 3 (Left panel). The unit of $B$ is normalized by the strength of magnetic field in the present-day sun, which is roughly of order $10^4$ Gauss in the solar tachocline and 3 Gauss in the atmosphere}. Fig. 3 exhibits notably different behaviour of $|B|$ in two different rotation rate regimes. For slow rotation rates, we can clearly see the increasing behavior of $|B|$ with rotation rate which attains a maximum value at $\Omega \approx 5.8$. {For $\Omega \in [1.17,5]$, the scaling of $|B|$ with respect to $\Omega$ is found to vary between 2.73 to 0.36. We observe an average scaling of 1.47 for $\Omega \in [1.25,2]$ which is close to the observed scaling of $1.38 \pm 0.14$ (Vidotto et al 2014). We note that $|A|$ also scales with $\Omega$ similarly to $|B|$}. Interestingly, there is a decrease in $|B|$ which continues up to $\Omega \simeq 12.5$. For $\Omega \ge 12.5$, that is, for very high rotation rates, $|B|$ fluctuates on a very rapid timescale, but with a cycle-averaged value, depicted in red,
that is essentially independent of $\Omega$. The rapid fluctuations in $|B|$ are due to the presence of two modes with different frequencies. The fluctuating behavior of $|B|$ with $\Omega$ can be seen in Fig. 3 (Right panel) for a small cut of $\Omega \in [23.30, 23.31]$.
 {Note how the system spends more time near
the top as opposed to the bottom, which explains why the cycle-averaged value
of $|B|$ (the red curve in the left panel) is higher than the simple average of
the cycle maxima and minima (the highs and lows of the blue curves).
(Observationally this would suggest that stars might be more likely to be
observed close to a peak of magnetic activity rather than a trough.)}
Furthermore, we notice a gap between the two different rotation rate regimes
in the region $\Omega \in [5.8, 12.5]$.
\begin{figure}[ht!]
\epsscale{1.2}
\plottwo{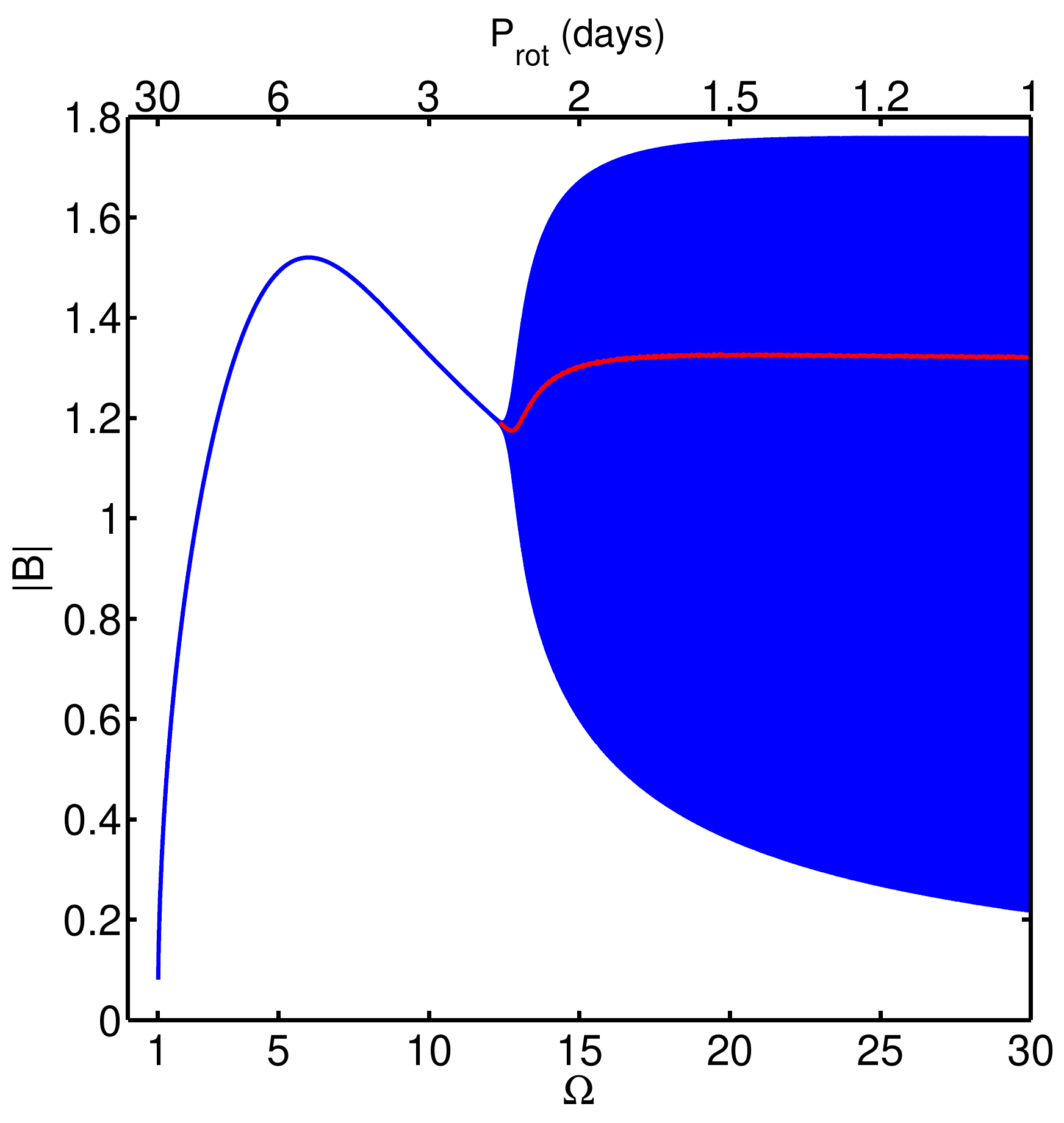}{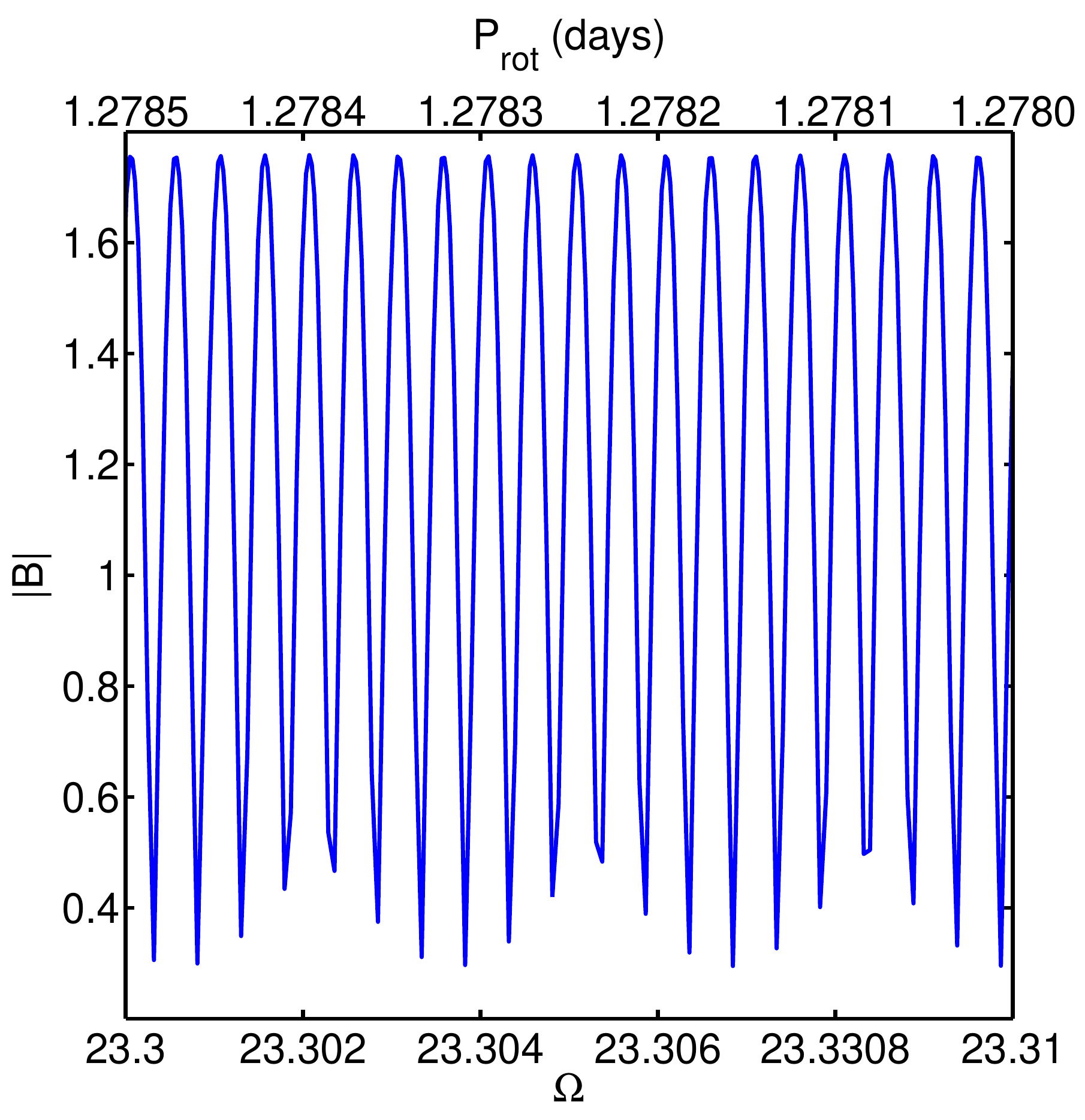}
\caption{Magnetic field strength $|B|$ is shown as a function of $\Omega$ (Left panel) with the cycle-average of $|B|$ depicted in red.  {We note that the value of $|B| \approx 0.1$ at $\Omega=1$ corresponds to about 3 G at the star's surface, whereas the rotation period $(P_{rot})$ in days is depicted on the top of the plot.} Also, fluctuating behavior $|B|$ for high rotation rate regime is shown for $\Omega \in [23.30, 23.31]$ (Right panel).}
\end{figure}

\subsection{Power spectra of $B$ and $\omega_{cyc}$ versus $\Omega$ relationship}

 {To understand how the rapid cycles in $|B|$ gradually evolve as
$\Omega$ spins down, we divided the entire time series into discrete chunks of 0.0106 Gyrs, and performed a Fourier transform on each chunk separately.  The precise length of the individual sections is not important, the only requirements being that it should be long compared with the fast cycle time, but short compared with the gradual spindown evolution time.}
Fig. 4 shows Fourier
spectra for 8 such sections. It is notable that at earlier times shown in the first and second rows, there are main two peaks around $\omega \sim 10$ in the spectra, whereas at later times only one in the third and fourth rows, with the peaks furthermore
shifting to lower frequencies.
In particular, in the second and third rows, where time increases from {age 0.1460 Gyrs to 0.2831 Gyrs}, we find that peaks shifting gradually towards lower frequency as time increases.
This behavior continues until we reach time approximately {age 0.3253 Gyrs} beyond which the multiple peaks of frequency are found to diminish. This behavior of frequency can be seen in panel 7 of Fig. 4 for time {$\approx$ [0.3148, 0.3253] Gyrs} while for time { $\approx$  [0.3569,  0.3675] Gyrs} we find only a single peak of frequency (see Fig. 4 panel 8). The behavior of power spectra of $|B|$ clearly shows that the second peak of frequency vanishes as time increases, that is, as the rotation rate decreases.
We note that in addition to the main two peaks at $\omega \sim 10$ or $\omega <10$ that we discussed above, one or two more peaks are also observed at higher frequency $\omega \sim 20$ in the first and second rows.
These high frequency modes have much weaker power than the main peaks and are simply their subharmonics. In the following, we do not
discuss these modes and only focus
on the behaviour of the main peaks (e.g. the higher frequency modes are not shown in Fig. 5).

\begin{figure}[ht!]
\epsscale{1.0}
\plotone{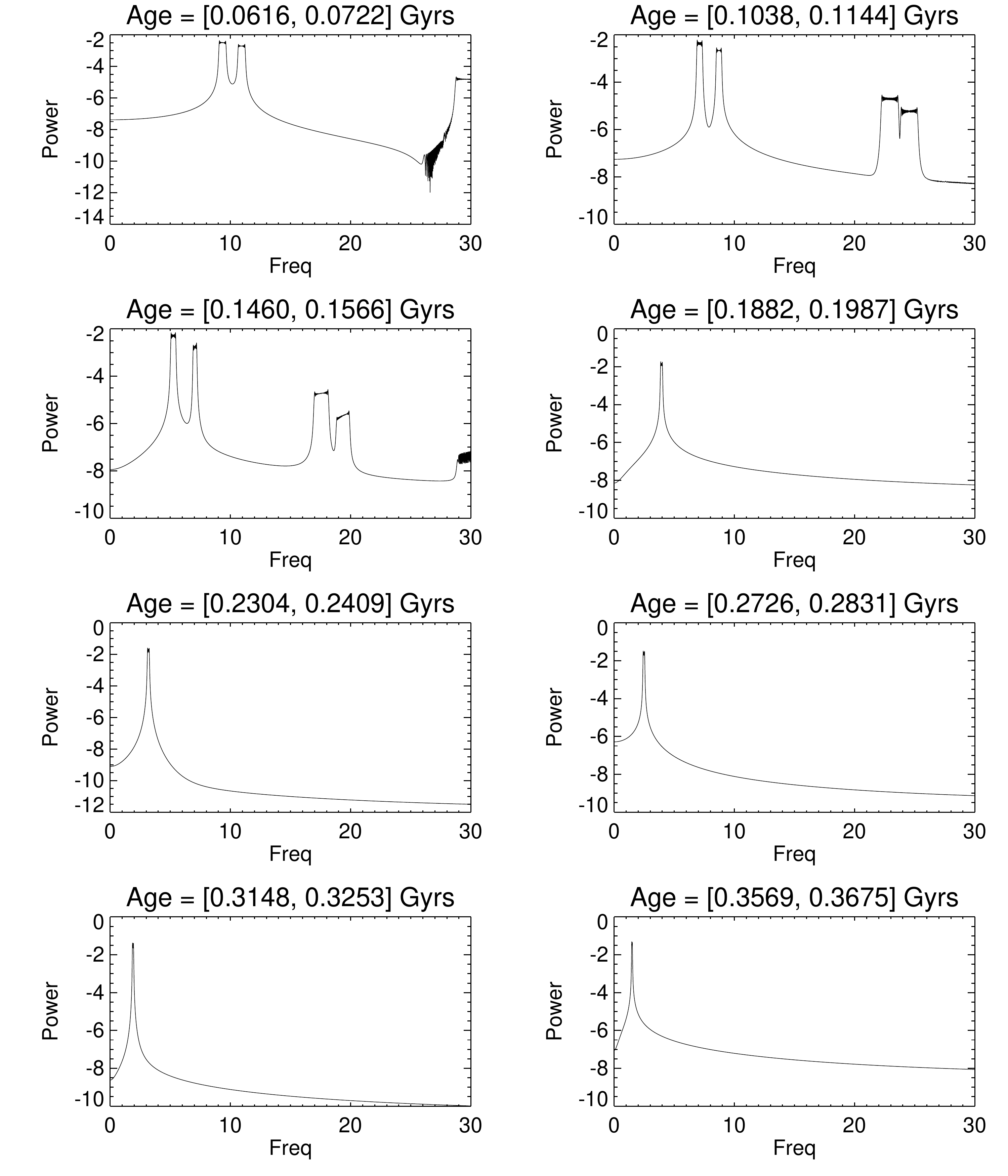}
\caption{Power spectra of $|B|$ for 8 distinct intervals in time, indicated by the numbers at the top of each panel.}
\end{figure}

 {The gradual transitions in the spectra of $|B|$ are further
illustrated in Fig. 5, showing so-called short-time Fourier
transforms (STFT). In this technique the signal is again
divided into short chunks, but these now overlap, essentially
forming a moving window, and hence giving an overview of the
continuous evolution of frequencies and amplitudes.} Using
this method, the most pronounced frequency of $|B|$ is
obtained in Fig. 5 (Left panel) where high to low intensity of frequency is illustrated via bright red to dark blue colors as shown in color map.
For early time $t < 0.3253 Gyrs$,  we observe two curves of frequency of maximum intensity $\omega_{cyc}$ (depicted in red) with {age in Gyrs}. {Lower curve} has larger amplitude of frequency than the {upper}  curve.
The existence of these two curves is the manifestation of complex time behaviour of fast rotators and is
reminiscent of the complexity of magnetic topology for active branch stars, discussed in recent papers (e.g. Matt et al 2015).
Both upper and lower curves show that the  frequency of maximum intensity decreases with age rapidly until $t \sim 0.3253 Gyrs$
when the upper curve disappears while the lower curve exhibits the change in the behaviour. This single curve for $t > 0.3253 Gyrs$
is interpreted as inactive branch. \\

In order to investigate further, we examine the scaling of $\omega_{cyc}$
by showing the behavior of frequency of maximum intensity $\omega_{cyc}$ against rotation rate $\Omega$ in the right panel in Fig. 5.
Again, we notice that for high rotation rate we have two curves of frequency for maximum intensity, whereas for slow rotation rate we have only one single curve. We use power law relationship, that is, $\omega_{cyc} \propto \Omega^{p}$ with a  power-law index $p$ to obtain the scaling.
\begin{figure}[ht!]
\epsscale{1.2}
\plottwo{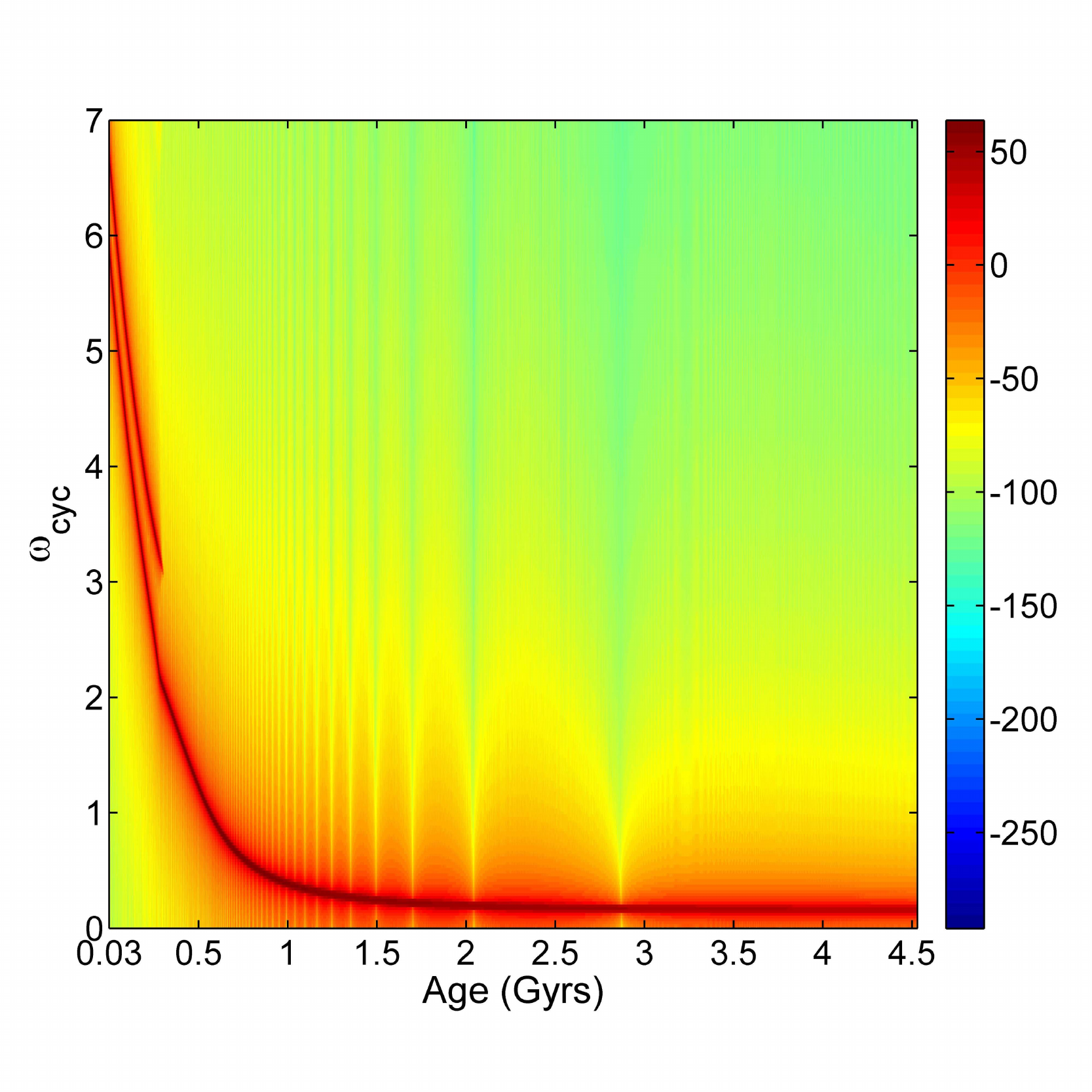}{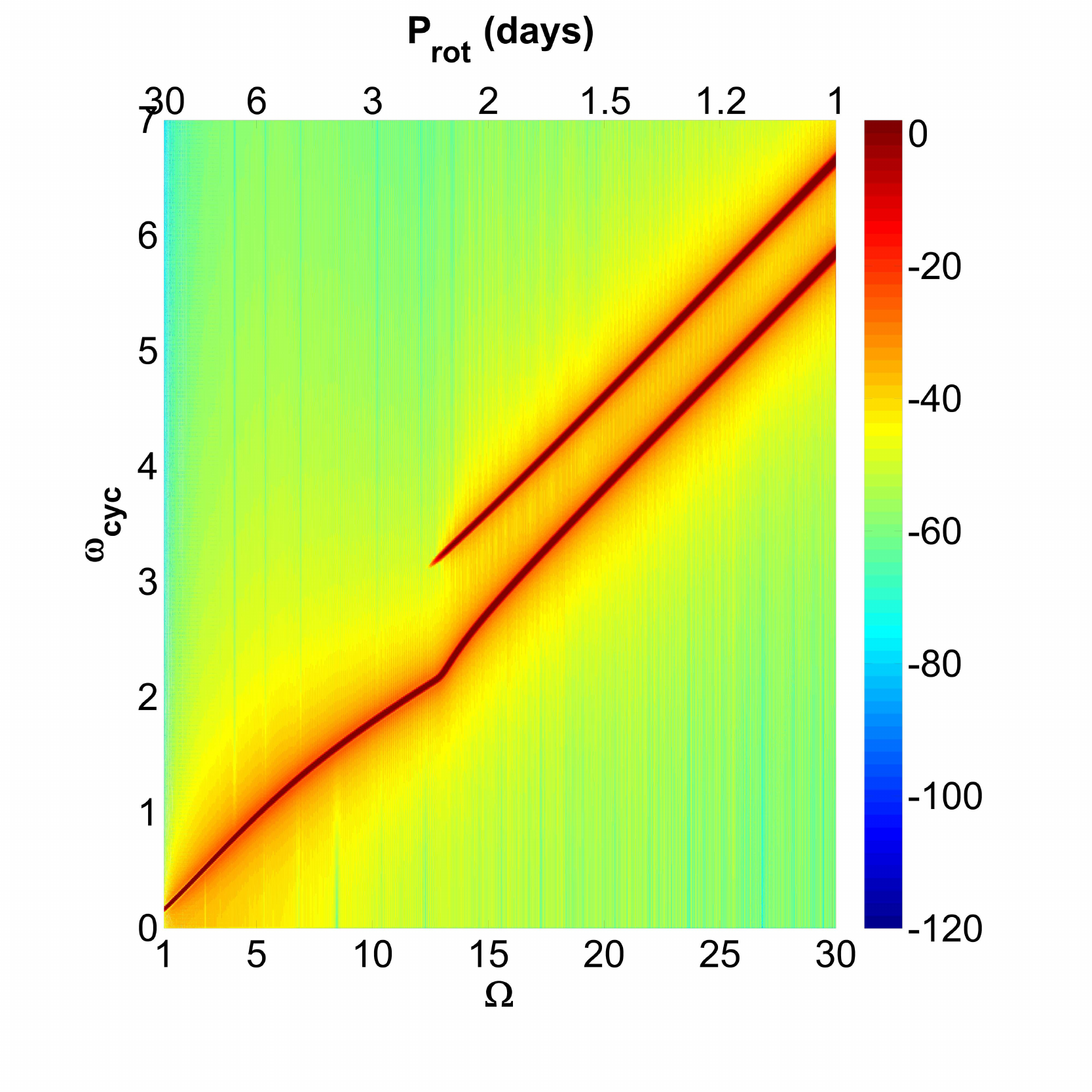}
\caption{Behavior of frequency of maximum intensity $\omega_{cyc}$, as a function of {age in Gyrs} and rotation rate
can be seen in left panel and right panel,
respectively. Here, bright red to dark blue colors represent high to low intensity of frequency. \label{fig2}}
\end{figure}
For the upper curve, we find the value of $p \sim 0.83$ for stars with rotation rate $12.8 \leq \Omega \leq 30$.
On the other hand, scaling exponent $p$ of the lower curve varies with rotation rate, as shown in Table 2.
Interestingly, for  fast rotator with $\Omega>12$, an average value of $p \sim 0.9$, which is close to the observational value
for active branch stars {(Saar \& Brandenburg 2001)}; for slow rotators, solar-like stars with rotation rate  in the range
$[1.17,3.5]$ has $p \sim 1.16$, in good agreement with observed scaling exponent for solar-type stars lying on inactive branch {(Saar \& Brandenburg 2001)}.\\

\begin{table}[ht!]
\caption{Power law exponent $p$ for the lower curve in  $\omega_{cyc} \propto \Omega^{p}$ at  different rotation periods.}              
\label{table:2}      
\centering
  \begin{tabular}{c c c }
    \hline
       \hline
     $p$ & $\Omega$ & $P_{rot}(days)$ \\ \hline
     1.16& $\Omega \in [1.17,3.5]$  & 25.65 -8.7 \\
     0.98& $\Omega \in [3.5,6]$ & 8.7 - 5 \\
    0.80& $\Omega \in [6,13]$ & 5 - 2.30\\
     1.06& $\Omega \in [16,30]$ & 1.88 - 1\\

    \hline
  \end{tabular}

\end{table}


\subsection{Total shear versus $\Omega$ relationship}

In our dimensionless units, the total shear is given by $1+w_{0}$. Fig. 6 shows how this total shear changes with rotation rate $\Omega$. As $\Omega$ increases from $\Omega =1$, the total shear is seen to decrease by $90\%$ from $1$ to $0.1$ with increasing $\Omega$. This reduction in total shear results from the effect of magnetic back-reaction on the shear. The saturation of the total shear for high rotation indicates that the dynamo efficiency is not saturated beyond certain rotation rate. After taking the minimum value around $\Omega =12.5$, the total shear increases with $\Omega$ in a small interval $\Omega\in[12.5,17]$ and then remains almost constant for high rotation rate $\Omega\ge17$. It is important to note that the apparently broad band of the total shear for $\Omega\ge12.5$ in Fig. 5 is due to the two different modes with different frequencies existing in this interval. The inset in Fig. 6 shows the total shear for a small range of $\Omega \in [29.82, 29.84]$ to highlight the fluctuation in total shear due to two modes.
Finally, $\Omega =12.5$, where the total shear takes its minimum value is related to very rapid transition in rotational evolution and is related to the V-P gap discussed later.
\begin{figure}[ht!]
\epsscale{0.6}
\plotone{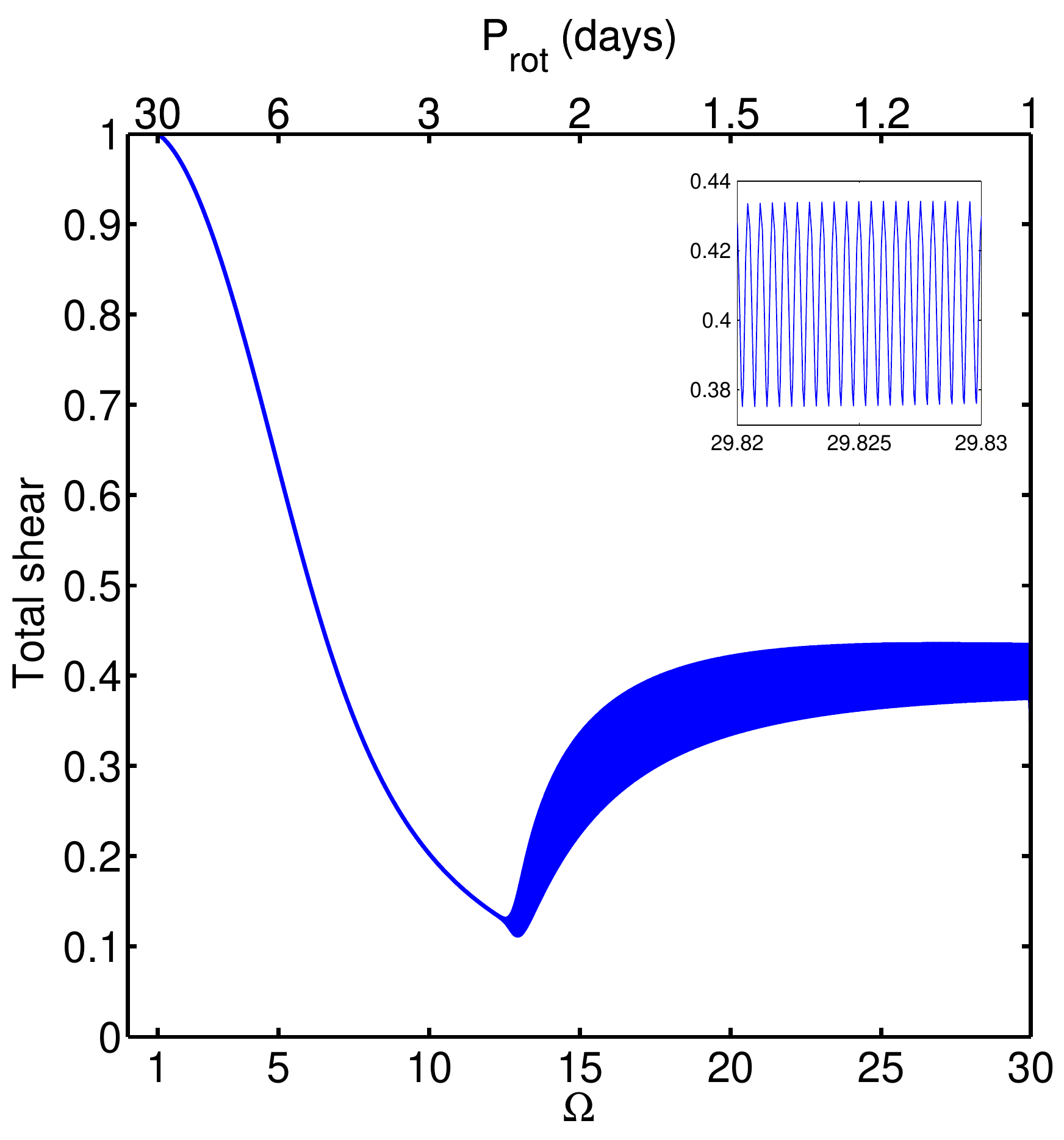}
\caption{Total shear $1+w_{0}$ as a function of rotation rate $\Omega$.}
\end{figure}

\subsection{$|B|$ versus Age Relationship}

In Fig. 7 magnetic field strength is shown as a function of age. The magnetic field strength $|B|$ is observed to maintain almost the same mean value fluctuating with finite amplitude for very young fast rotating stars of age up to 325 Myrs. This fluctuation is due to the presence of two different modes as discussed later. The magnetic activity is seen to increase with age in the range $\in {[325,502]}$ Myrs as $|B| \sim t^{s}$ with a power law exponent $s=0.53$ after which the magnetic activity remains almost constant in the age interval $\in{[508,551]}$ Myrs. Beyond this value the magnetic activity decreases very rapidly with increasing age. We find that power law exponent $s$ varies with different values for stars with different ages and are provided in Table 3.
{
Finally, we note that our results suggest that the fraction of poloidal vs toroidal flux fluctuate for fast rotators,
consistent with complex magnetic topology (e.g. Matt et al 2015), while it takes a constant value
for slow rotators. The mean value of this ratio does not change significantly over time.
}

\begin{figure}[ht!]
\epsscale{0.6}
\plotone{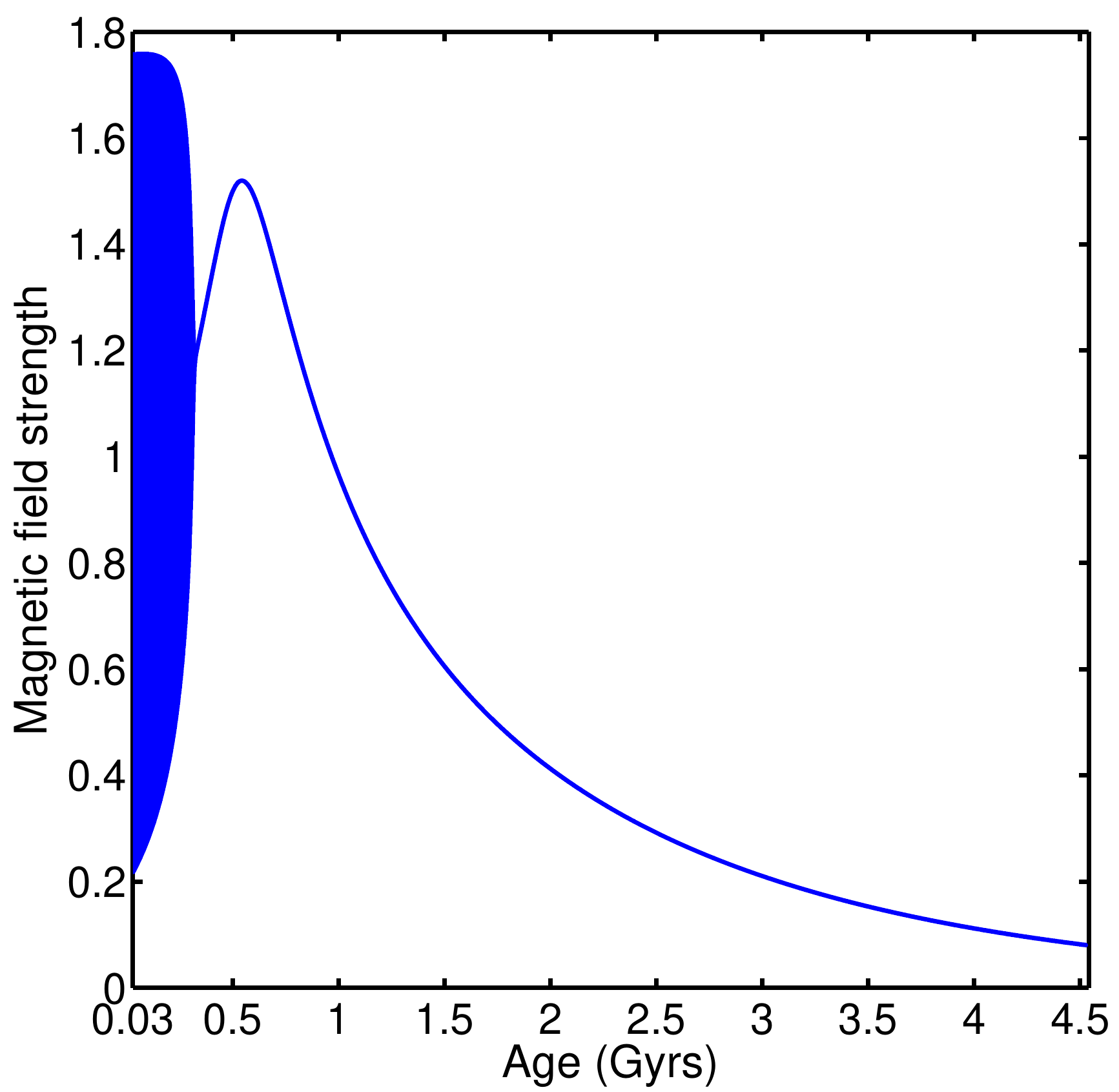}
\caption{Magnetic field strength $|B|$ as a function of age.}
\end{figure}
\begin{table*}[ht!]
\caption{Power law exponent $s$ for magnetic activity with age $t$ of stars $\in [1.066,4.5]$ Gyrs}
\label{table:1}      
\centering
  \begin{tabular}{c c}
    \hline
       \hline
      $s$ & Age (Gyrs)  \\     \hline
     -0.61&  $t\in {[0.5929, 0.7336]}$\\
     -0.97&  $t\in {[0.7336, 1.085]}$  \\
     -1.13&  $t\in {[1.085, 1,437]}$ \\
     -1.25&  $t\in {[1.437, 1.788]}$ \\
     -1.40&  $t\in {[1.788, 2.139]}$ \\
     {-1.64}&  $t\in {[2.139, 2.843]}$ \\
     {-2.00}&  $t\in {[2.843, 3.54]}$ \\
     {-2.50}&  $t\in {[3.54, 4.5]}$ \\

    \hline
  \end{tabular}
\end{table*}

 \subsection{Timescale for spindown}
 \begin{figure}[ht!]
\epsscale{1.0}
\plottwo{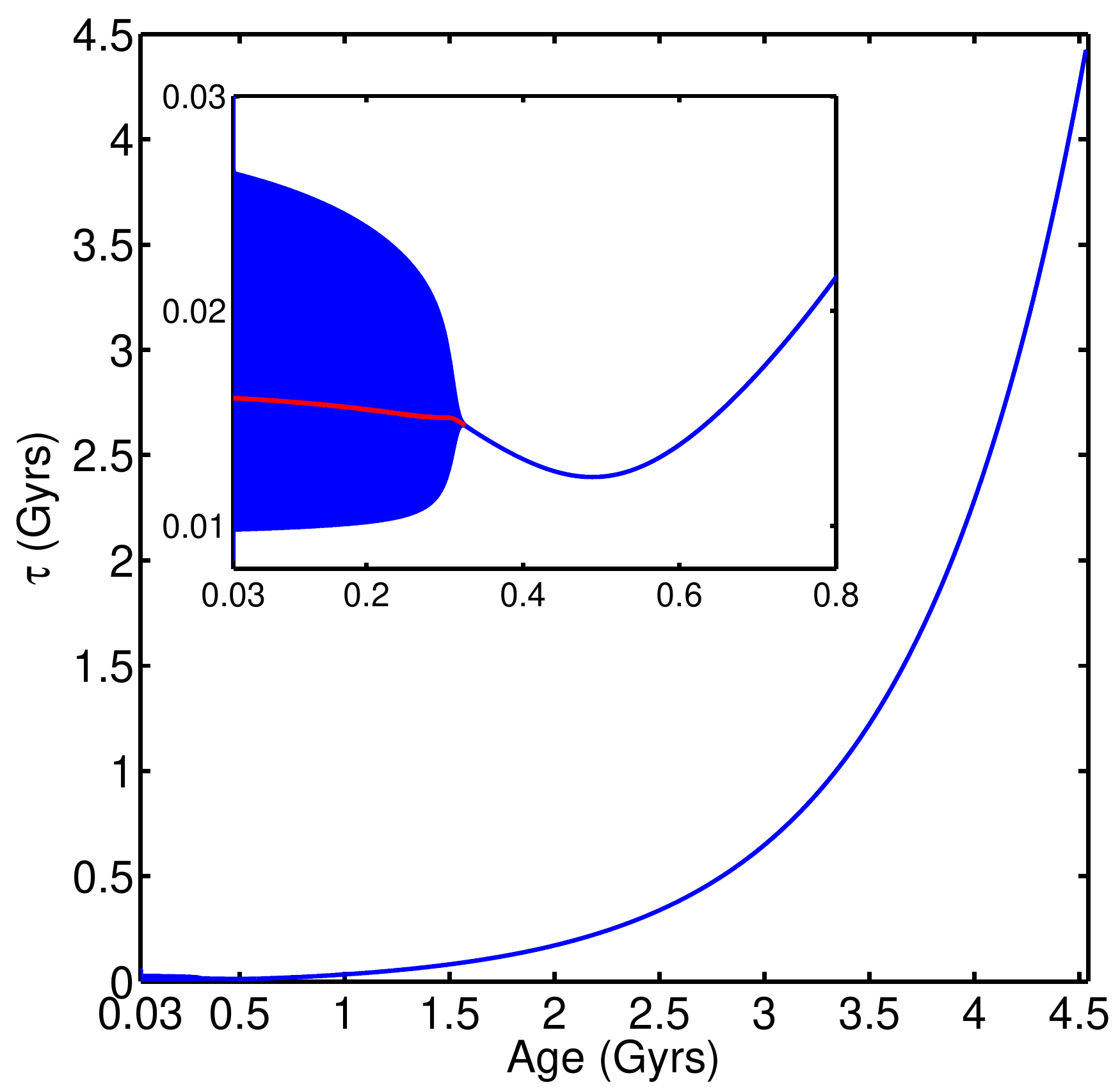}{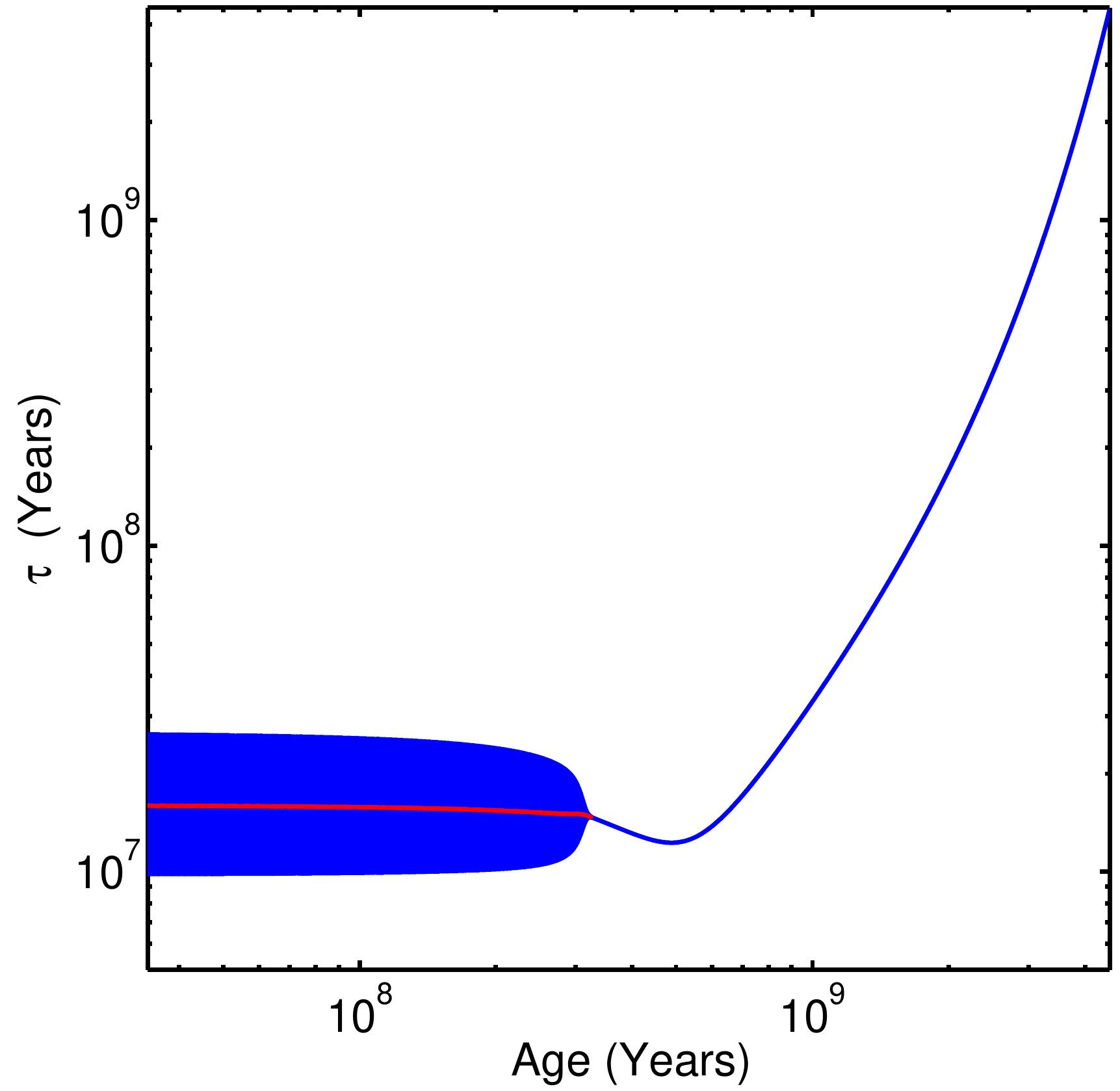}
\caption{Left panel shows the spindown timescale $\tau$ as a function of age in Gyrs in linear scale while right panel
 shows $\tau$ as a function of age in years in log-scale. The oscillations in $\tau$ are caused by the fluctuations in $\dot\Omega$ previously seen in Fig. 1 (Right panel).}
\end{figure}
In order to quantify the timescale of spindown, we compute the characteristic spindown time $\tau = |\Omega/\dot \Omega|$ by using
{Eq. 6}
as we evolve the system, and then show the suitable averaged value in Fig. 8 using linear and log scales.
{The inset in Fig. 8 (left panel) shows the zoomed in view of $\tau$ for very fast rotating stars.}
 Here, red depicts the mean value$\footnote{\label{ave}This mean value is obtained over the $\frac{1}{420}$ fraction of the interval.}$  of timescale over time. Clearly, we observe a spindown timescale of {15.96 Myrs for very young rapidly
rotating stars of ages from 30 Myrs which decreases very slowly up to age 315 Myrs. Beyond this, the spindown timescale is observed to decrease rapidly with increasing age for a short interval $[315,493.9]$ Myrs}. This decline in spindown time reaches a minimum of approximately {122 Myrs for age 493.9 Myrs}. After this, the spindown timescale starts increasing with age of the stars. Specifically, the spindown timescale increases linearly for solar-type stars with ages approximately 4.5 Gyrs. The shortest spindown timescale is obtained in the region {$[315,632]$ }Myrs ($\Omega \in [5.8, 12.5]$) noted previously, and interestingly corresponds to the V-P gap, the transition region between fast and slow rotators. That is, this is the region where the star suddenly jumps from active to inactive branch staying in this intermediate region for a short time only due to the fast spindown. To summarise, our results show that spindown time for fast rotating stars in that region is shorter than the spindown time for slow rotating stars while the spindown timescale for stars in the transition region is even much shorter than the spindown timescale for fast rotating stars. These results are in good agreement with observations for spindown timescale (Barnes, 2003).

\subsection{Summary of results}

{Our dynamical model of spindown coupled to the evolution of
magnetic fields successfully reproduced: (i) the basic $\Omega$
versus age relationship, (ii) the relationships of $|B|$,
$\omega_{cyc}$ and total shear $1+w_0$ versus $\Omega$, and
(iii) magnetic activity and spindown timescales with age. All
three items are consistent with observations for the spindown
process for fast and slow rotating stars, and provide a natural
explanation for the V-P gap associated with the abrupt
transition of stars from active ($A$) to inactive ($I$)
branches, which is an important unresolved issue.}\\

\section{Conclusions}
{The evolution of magnetic fields and rotation rate is a self-regulated process through the direct interaction between
large-scale shear flow and magnetic field and the indirect interaction by various (nonlinear) feedback mechanisms through small-scale fields.
In particular, the generation of magnetic field and spindown are closely inter-linked processes since the generation of magnetic field depends on rotation of stars and thus spindown
while spindown process crucially depends upon the magnetic field (e.g. generation, destruction) and differential rotation.
In this paper, we have proposed a dynamical model of  spindown to understand self-regulation of magnetic fields and rotation over the spindown time scale, which for the first time evolves  magnetic field and rotation rate at the same time taking into account various mutual interactions.
Despite being a simple parameterized model our model successfully reproduces the observations for
spindown of stars which would otherwise be impossible in a  more complete model (e.g. 3D MHD).
In particular, we have found exponential spindown, saturation of magnetic field strength and power
law dependence of frequency of magnetic fields of active and inactive branches  for rapidly rotating stars. For slow rotators, we obtained power law spindown, linear
scaling of magnetic field strength and power law relationship of $\omega_{cyc} $ on $\Omega$ with power law scaling for inactive branch.
The transition from fast to slow rotating stars is quantitatively shown to occur very rapidly, thereby providing a natural
 explanation for the V-P gap.}
{
In future, interesting extension of our model would include the coupling of our models to the evolution of mass-loss
(e.g. Garraffo 2015) and detailed modelling of  slow rotating stars  (Saders et al 2016) and transient Sun (Metcalfe 2016).\\
}

\acknowledgments
We thank Dr M Miesch for valuable discussion.

\end{document}